\input{style/arxiv-general.cfg}
\documentclass[aoas,MSNbibl,nameyear,seceqn,rotating,dvips]{arximspdf}
\makeatletter
   \@ifpackageloaded{graphicx}{}{\usepackage{graphicx}}
\makeatother
\usepackage{dcolumn}
\usepackage{graphicx}

\doi{10.1214/14-AOAS791} 
\volume{9}
\issue{1}
\pubyear{2015}
\firstpage{275}
\lastpage{299}
\docsubty{FLA}

\makeatletter
\newcolumntype{d}[1]{D{.}{.}{#1}}
\newcommand{\cov}{\operatorname{cov}}
\newcommand{\rrvert}{\vert}
\renewcommand{\mid}{|}
\makeatother
\setattribute{copyright}{owner}{\textup{In the Public Domain}}

\begin{document}
\begin{frontmatter}

\title{Mixed model and estimating equation approaches for zero
inflation in clustered binary response data with application to a
dating violence study\thanksref{T1}}
\runtitle{Zero inflation in clustered binary response data}

\begin{aug}
\author[A]{\fnms{Kara A.}~\snm{Fulton}},
\author[A]{\fnms{Danping}~\snm{Liu}\corref{}\ead[label=e2]{danping.liu@nih.gov}},
\author[A]{\fnms{Denise L.}~\snm{Haynie}}
\and
\author[A]{\fnms{Paul S.}~\snm{Albert}}
\runauthor{Fulton, Liu, Haynie and Albert}
\affiliation{Eunice Kennedy Shriver National Institute of Child
Health and Human Development}
\address[A]{Biostatistics and Bioinformatics Branch\\
\quad and Health Behavior Branch\\
Division of Intramural Population Health Research\\
Eunice Kennedy Shriver National Institute of Child Health\\
\quad and Human Development\\
Bethesda, Maryland 20852\\
USA\\
\printead{e2}}
\end{aug}
%
\thankstext{T1}{Supported by the Intramural
Research Program of the National Institute of Health
(NIH), Eunice Kennedy Shriver National
Institute of Child Health and Human Development
(NICHD). The NEXT Generation Health Study was
supported in part by the Intramural Research Program
of the NIH, Eunice Kennedy Shriver National Institute
of Child Health and Human Development (NICHD),
and the National Heart, Lung and Blood Institute
(NHLBI), the National Institute on Alcohol Abuse and Alcoholism (NIAAA), and Maternal and Child Health
Bureau (MCHB) of the Health Resources and Services Administration (HRSA), with supplemental support
from the National Institute on Drug Abuse (NIDA).}

\received{\smonth{7} \syear{2013}}
\revised{\smonth{8} \syear{2014}}

%
\begin{abstract}
The NEXT Generation Health study investigates the dating violence of adolescents
using a survey questionnaire. Each student is asked to affirm or deny multiple
instances of violence in his/her dating relationship. There is,
however, evidence suggesting that students not in a relationship
responded to the survey, resulting in excessive zeros in the responses.
This paper proposes likelihood-based and estimating equation approaches
to analyze the zero-inflated clustered binary response data. We adopt a
mixed model method to account for the cluster effect, and the model
parameters are estimated using a maximum-likelihood (ML) approach that
requires a
Gaussian--Hermite quadrature (GHQ) approximation for implementation.
Since an incorrect assumption on the random effects distribution may
bias the results, we construct generalized estimating equations (GEE)
that do not require the correct specification of within-cluster
correlation. In a series of simulation studies, we examine the
performance of ML and GEE methods in terms of their bias, efficiency
and robustness. We illustrate the importance of properly accounting for
this zero inflation by reanalyzing the NEXT data where this issue has
previously been ignored.
\end{abstract}

%
\end{frontmatter}

\section{Introduction}\label{sec1}

In public health studies, clustered or longitudinal binary responses
may be collected on a group of individuals where only a subset of
these individuals are susceptible to having a positive response. For
example, questionnaires may ask teenagers who are dating to answer
a series of questions about dating violence. As in the NEXT
Generation Health Study, a larger proportion of all zero responses are observed
than would occur by chance; presumably many\vadjust{\goodbreak} individuals who are not
dating filled in all zeros on the questionnaire (also known as
``structural zeros'').
While there may be alternative reasons for structural zeros, for
example, participants giving socially desirable responses, we believe
this accounts for only a small fraction of zero inflation. Interest is
in making
inference about the correlated binary responses for those who are
susceptible (i.e., inference about dating violence among
individuals who were dating).

There is an extensive literature on zero-inflated Poisson and binomial
models [\citet{Lam92}; \citet{Hal00}] that provide early references,
along with more recent work on zero-inflated ordinal data [\citet{KelAnd08}] and zero-inflated sum score data with randomized
responses [\citet{Cruetal08}]. \citet{MinAgr02} reviewed
various statistical models incorporating zero inflation in both discrete
and continuous outcomes for cross-sectional data. \citet{DioDioDup11}
discussed cross-sectional binary regression with zero inflation, and
proved the model
identifiability when at least one covariate is continuous. \citet{Hal00} first
considered longitudinal or clustered data with zero-inflated binomial
or Poisson outcomes. They incorporated a random effect structure to
model the within-subject correlation and proposed an EM algorithm
to estimate the parameters. \citet{HalZha04} extended the work
of \citet{Hal00} by proposing a generalized estimation
equation (GEE) approach to model several zero-inflated distributions
in a longitudinal setting. \citet{MinAgr05} presented a Hurdle
model with random effects for repeated measures of zero-inflated count
data. There has been no work, however, on zero-inflated clustered
binary data.

A component of the NEXT Generation Health Study examines the prevalence
and correlates of dating violence among 2787 tenth-grade students,
following them over seven years. Dating violence is common among adolescents,
may impact adolescent expectations regarding adult intimate relationships
[\citet{Col03}], and has been found to be associated with increased
risk of depression and engagement in high-risk behaviors
[\citet{AckEisNeu07} and \citet{ExnEckRot13}]. Thus, dating violence among adolescents
merits interest from both developmental and public health perspectives
[\citet{OffBuc11}].

%
\begin{figure}

\includegraphics{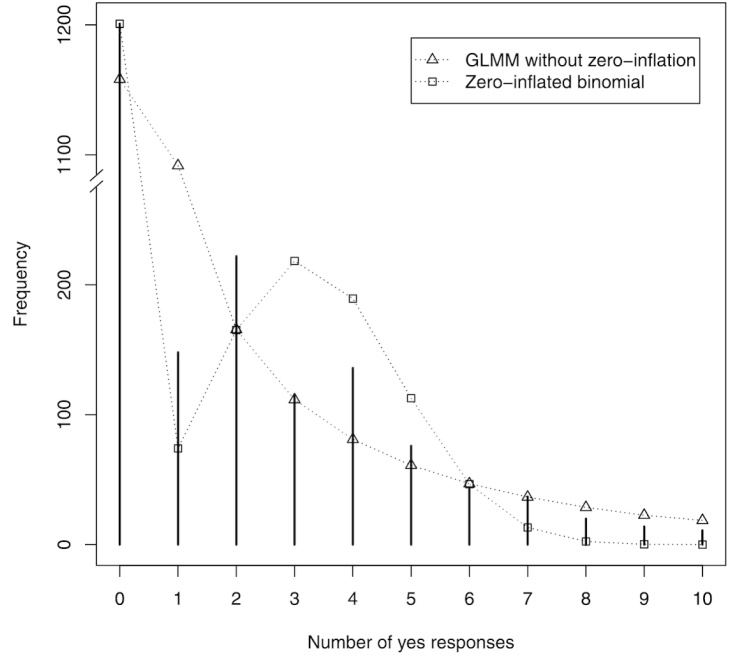}

\caption{Distribution of subjects' responses to five dating violence
victimization questions and the fitted probabilities using a
zero-inflated binomial model (black squares).}\label{fig1}
\end{figure}

Investigators involved in the NEXT study are primarily interested
in identifying the risk factors associated with dating violence. \citet{Hayetal13}
found a relationship between high-risk behaviors
(i.e., depressive symptoms, alcohol use, smoking and drug use), gender
and the prevalence of dating violence victimization. A total of 10
questions were asked about dating violence. Five of the questions
were on dating violence victimization: did your partner (1) insult
you in front of others, (2) swear at you, (3) threaten you, (4) push
or shove you, or (5) throw anything that could hurt you; the other
five were similar questions on perpetration: did you (1) insult your
partner in front of others, (2) swear at your partner, (3) threaten
your partner, (4) push or shove your partner, or (5) throw anything
that could hurt your partner? As illustrated in Figure~\ref{fig1},
the distribution of the number of ``yes'' responses is clumped at
zero. When we fit the frequencies with a zero-inflated binomial distribution,
the zero-inflation probability is estimated to be about 58\%.\vadjust{\goodbreak} The
binomial distribution yields a poor fit to the frequencies for two reasons.
First, the prevalence of ``yes'' responses is unequal across different
questions; second, the responses from the same
subject are correlated.
But this only serves as an intuitive visualization of zero inflation.
One can argue that the clump of zeros might be due to the high correlation
of the binary responses within the same subject; and, therefore, we also
fit the generalized linear mixed model (GLMM) and plot the fitted
frequencies in Figure~\ref{fig1}. GLMM attempts to fit the spike
at 0, and hence tends to overestimate the within-subject correlation.
In this paper, we hope to explore whether zero inflation exists while
allowing for the cluster effects. 
We propose maximum-likelihood (ML) and GEE approaches to simultaneously
account for the zero inflation and clustering in the multiple binary
responses. The major difference between our work and the previous
work is that \citet{Hal00}, \citet{HalZha04}, and \citet{MinAgr05} all considered the zero inflation at the \textquotedblleft{}observation
level,'' while in our paper the zero inflation is
at the \textquotedblleft{}subject level'' (meaning
that with a structural zero, all the binary responses from a subject
are zero). For our dating violence example, subjects have
all zero responses because they are not susceptible to the condition
(e.g., in a relationship). The proposed methods are evaluated and
compared in simulation studies. We then reexamine the relationships
between high-risk behaviors and dating violence among teenagers using the
proposed analysis strategy accounting
for zero inflation.

In Section~\ref{sec2} we present both maximum-likelihood and GEE approaches
for parameter estimation. Section~\ref{sec3} discusses the identifiability
of the proposed model and proposes a likelihood ratio test for zero inflation.
Simulation study results are presented in Section~\ref{sec4}. The NEXT dating
violence data is analyzed in Section~\ref{sec5}, and a discussion
follows in Section~\ref{sec6}.

\section{Method}\label{sec2}

Let $\mathbf{Y}_{i}=(Y_{i1},\ldots,Y_{iJ})'$ be the multivariate
binary outcome for subject $i$ ($i=1,\ldots,N$), and $\mathbf
{X}_{i}=(\mathbf{X}_{i1},\ldots,\mathbf{X}_{iJ})'$
be the corresponding matrix of covariates. Let $Z_{i}$ be the latent
class, so that $\mathbf{Y}_{i}$ always takes the value of $\mathbf{0}$
(structural zero) if $Z_{i}=0$, and $\mathbf{Y}_{i}$
follows a multivariate binary distribution with density $f(\mathbf
{Y}_{i};\bolds{\theta})$
if $Z_{i}=1$, where $\bolds{\theta}$ is a vector of parameters.
We suppress the subscript~$i$ when there is no confusion. Let $p=\Pr(Z=1)$
be the prevalence of the latent class~1. In our example, $Z_{i}=1$ indicates
that subject $i$ is susceptible to the possibility of dating violence
(i.e., potential of answering the dating violence questions in a positive
fashion), while $Z_{i}=0$ indicates that the subject is not susceptible.

\subsection{Maximum-likelihood estimation}\label{sec2.1}

If both $\mathbf{Y}$ and $Z$ are observed, the individual contribution
to the full data likelihood is
\[
L^{F}(\mathbf{Y},Z;\bolds{\theta})= \bigl\{ I(\mathbf{Y}=
\mathbf{0}) (1-p ) \bigr\} ^{1-Z} \bigl\{ f(\mathbf{Y};
\bolds{\theta})p \bigr\} ^{Z}.
\]
The observed likelihood of $\mathbf{Y}$ is then given by
\begin{eqnarray*}
L (\mathbf{Y};\bolds{\theta} ) & = & L^{F}(
\mathbf{Y},Z=0;\bolds{\theta})+L^{F}(\mathbf{Y},Z=1;
\bolds{\theta})
\\
& = & I(\mathbf{Y}=\mathbf{0}) (1-p )+f(\mathbf{Y};\bolds{\theta})p.
\end{eqnarray*}
Here we assume that the zero-inflation probability $p$ is the same
across all the subjects in the sample. This could easily be extended
to allow $p$ to depend on covariates, for example, with a logistic regression
model. We use a generalized linear mixed effects model (GLMM) to describe
the multivariate distribution, $f(\mathbf{Y};\bolds{\theta})$:
\[
g \bigl\{ \pi_{ij}(\mathbf{b}_{i}) \bigr\} =
\mathbf{X}_{ij}'\bolds{\gamma}+
\mathbf{Z}_{ij}'\mathbf{b}_{i},
\label{eq0-glmm}
\]
where $\pi_{ij}(\mathbf{b}_{i})=\Pr(Y_{ij}=1\mid\mathbf{X}_{ij},\mathbf
{b}_{i})$,
$\mathbf{b}_{i}$ is the vector of random effects following the
multivariate normal distribution $\mathrm{MVN}(0,\Delta)$, $\mathbf{Z}_{ij}$
is the design matrix of the random effects, and $g$ is the known
link function. The parameter vector $\bolds{\theta}$ consists
of the parameter of interest $\bolds{\gamma}$ and the nuisance
parameters in the variance component $\Delta$. Assume $Y_{ij}$'s
are mutually independent given $\mathbf{X}_{ij}$ and $\mathbf{b}_{i}$,
and let $p(\mathbf{b}_{i};\Delta)$ be the probability density
function of $\mathbf{b}_{i}$. Then the likelihood for subject
$i$ becomes
\begin{eqnarray*}
L (\mathbf{Y}_{i};\bolds{\theta} ) &=& I(\mathbf{Y}_{i}=
\mathbf{0}) (1-p )
\\
&&{} +p\int\Biggl\{ \prod_{j=1}^{J}
\pi_{ij}(\mathbf{b}_{i})^{Y_{ij}}\bigl(1-
\pi_{ij}(\mathbf{b}_{i})\bigr)^{1-Y_{ij}} \Biggr\} p(
\mathbf{b}_{i};\Delta)\,d\mathbf{b}_{i}.
\end{eqnarray*}
The integral with respect to the random effects can be approximated
by Gaussian--Hermite quadrature as
\begin{eqnarray*}
&& \int\Biggl\{ \prod_{j=1}^{J}
\pi_{ij}(\mathbf{b}_{i})^{Y_{ij}}\bigl(1-
\pi_{ij}(\mathbf{b}_{i})\bigr)^{1-Y_{ij}} \Biggr\} p(
\mathbf{b}_{i};\Delta)\,d\mathbf{b}_{i}
\\
&&\qquad \approx\sum
_{q=1}^{Q} \Biggl\{ w_{q}\times
\prod_{j=1}^{J}\pi_{ij}(
\mathbf{b}_{i,q})^{Y_{ij}}\bigl(1-\pi_{ij}(
\mathbf{b}_{i,q})\bigr)^{1-Y_{ij}} \Biggr\} ,
\end{eqnarray*}
where $\mathbf{b}_{i,q}$ is the $q$th quadrature grid point
and $w_{q}$ is the associated weight [\citet{AbrSte72}].

The parameter estimation for $p$ and $\bolds{\theta}$ can be
found by maximizing the log-likelihood for all $N$ subjects, $\sum
_{i=1}^{N}\log L (\mathbf{Y}_{i};\bolds{\theta} )$.
The variance estimation is calculated from the inverse of the observed
information:
\[
\Biggl(-\sum_{i=1}^{N}\frac{\partial^{2}}{\partial(p,\bolds{\theta})^{2}}
\log L (\mathbf{Y}_{i};\bolds{\theta} ) \Biggr)^{-1},
\]
and can be implemented by the \texttt{optim} function in \texttt{R} [\citet{autokey23}].

\subsection{Generalized estimating equations (GEE)}\label{sec2.2}

Likelihood-based inference makes full distributional assumptions
on $\mathbf{Y}\mid Z=1$. When these assumptions are correct, the
estimator gains efficiency; otherwise, classical inference has poor
statistical properties. We explore the estimating
equations approach [\citet{LiaZeg86}] that only specifies a structure
for the conditional mean $E (\mathbf{Y}\mid Z=1,\mathbf{X}_{i} )$.
Suppose
%
\begin{equation}
\bolds{\mu}_{i}^{Z}=E (\mathbf{Y}_{i}
\mid Z_{i}=1,\mathbf{X}_{i} )=g(\mathbf{X}_{i}
\bolds{\beta}),\label{eq0-mean}
\end{equation}
where $g$ is the known link function and $\bolds{\beta}$ is
the regression coefficients of interest. Unconditional on $Z_{i}$,
the ``marginal'' mean of $\mathbf{Y}_{i}$
is given by
\begin{eqnarray*}
\bolds{\mu}_{i}^{M} & = & E (\mathbf{Y}_{i}
\mid\mathbf{X}_{i} )=g(\mathbf{X}_{i}\bolds{\beta})\times\Pr
(Z=1)+\mathbf{0}\times\Pr(Z=0)
\\
& = & pg(\mathbf{X}_{i}\bolds{\beta}).
\end{eqnarray*}
The estimating equations can then be written as
%
\begin{equation}
\sum_{i=1}^{N}D_{i}'V_{i}^{-1}
\bigl(\mathbf{Y}_{i}-\bolds{\mu}_{i}^{M}
\bigr)=0,\label{eq1-gee1}
\end{equation}
where $D_{i}=\frac{\partial\bolds{\mu}_{i}^{M}}{\partial(p,\bolds{\beta})}$ and
$V_{i}$ is the working covariance matrix for $\mathbf{Y}_{i}$
[\citet{LiaZeg86}]. We can decompose $V_{i}$ as
$A_{i}^{1/2}R_{i}A_{i}^{1/2}$
with $A_{i} $ being\vspace*{1pt} the diagonal matrix of the variance of $Y_{ij}$
[which is $\mu_{ij}^{M}(1-\bolds{\mu}_{ij}^{M})$] and $R_{i}$
being the working correlation matrix specified by some nuisance parameter
$\bolds{\eta}$.

If the mean model (\ref{eq0-mean}) is correct, the estimating equations
(\ref{eq1-gee1}) are always consistent regardless of the working
correlation, and choosing an approximately\vadjust{\goodbreak} correct working correlation
generally leads to improved efficiency. In the context of zero-inflated
regression, we propose two ways to specify the working correlation:
marginal and conditional specification. The marginal correlation directly
makes assumptions on $R_{i}$, which is similar to the standard GEE:
the \emph{marginal independent correlation} assumes $R_{i}^{\mathrm{MI}}=\mathbf
{I}_{J\times J}$,
the $J$-dimensional identity matrix; the \emph{marginal exchangeable
correlation} assumes that $R_{i}^{\mathrm{ME}}=(1-\alpha)\mathbf{I}_{J\times
J}+\alpha\mathbf{1}_{J\times J}$,
where $\mathbf{1}_{J\times J}$ is the $J\times J$ square matrix
of ones. We refer to these two different approaches as GEE-MI and
GEE-ME, respectively.

The conditional correlation exploits the zero-inflated structure
and utilizes the conditional covariance, $V_{i'}^{Z}=\cov(\mathbf
{Y}_{i}\mid Z_{i}=1)$,
to derive the unconditional covariance $\cov(\mathbf{Y}_{i})$. A similar
idea was first
used by \citet{HalZha04} to derive their GEE estimator for
observation-level zero inflation.
By the law of total covariance, for $j\neq j'$,
\begin{eqnarray*}
\cov(Y_{ij},Y_{ij'}) & = & E\bigl(\cov(Y_{ij},Y_{ij'}
\mid Z)\bigr)+\cov\bigl(E (Y_{ij}\mid Z ),E (Y_{ij'}\mid Z )
\bigr)
\\
& = & E\bigl(V_{i,jj'}^{Z}Z\bigr)+\cov\bigl(
\mu_{ij}^{Z}Z,\mu_{ij'}^{Z}Z\bigr)
\\
& = & V_{i,jj'}^{Z}p+\mu_{ij}^{Z}
\mu_{ij'}^{Z}p(1-p),
\end{eqnarray*}
where $V_{i,jj'}^{Z}$ is the $(j,j')$ element of
$V_{i'}^{Z}$, and $\mu_{ij}^{Z}$ is the $j$th element of
$\bolds{\mu}_{i}^{Z}$. The variance of $Y_{ij}$ is given by
$\operatorname{var}(Y_{ij})=\mu_{ij}^{M}(1-\bolds{\mu}_{i}^{M})$. The \emph{conditional
independence correlation} assumes that $V_{i,jj'}^{Z}=0$,
so the working correlation is $R_{i}^{\mathrm{CI}}$ with the $(j,j')$
element as
\[
R_{i,jj'}^{\mathrm{CI}}=\frac{\mu_{ij}^{Z}\mu_{ij'}^{Z}p(1-p)}{\sqrt{\mu
_{ij}^{Z}p(1-\mu_{ij}^{Z}p)\mu_{ij'}^{Z}p(1-\mu_{ij'}^{Z}p)}}.
\]
The \emph{conditional exchangeable correlation} assumes that
\[
V_{i,jj'}^{Z}=\alpha\sqrt{\mu_{ij}^{Z}\bigl(1-\mu_{ij}^{Z}\bigr)\mu
_{ij'}^{Z}\bigl(1-\mu_{ij'}^{Z}\bigr)},
\]
that is, a correlation of $\alpha$ between any $Y_{ij}$ and $Y_{ij'}$
\emph{given} $Z=1$. Therefore, the $(j,j')$ element of the
working correlation $R_{i}^{\mathrm{CE}}$ is
\[
R_{i,jj'}^{\mathrm{CE}}=\frac{\alpha p\sqrt{\mu_{ij}^{Z}(1-\mu_{ij}^{Z})\mu
_{ij'}^{Z}(1-\mu_{ij'}^{Z})}+\mu_{ij}^{Z}\mu_{ij'}^{Z}p(1-p)}{\sqrt{\mu
_{ij}^{Z}p(1-\mu_{ij}^{Z}p)\mu_{ij'}^{Z}p(1-\mu_{ij'}^{Z}p)}}.
\]
We refer to these conditional GEE approaches as GEE-CI and GEE-CE,
respectively.

Similar to the ordinary GEE, an unstructured working correlation can
be assumed that allows for distinct correlations for each pair of
outcomes. With the unstructured GEE, the marginal and conditional
specification of working correlation are equivalent, that is,
\[
R_{i,jj'}^{\mathrm{UN}}=\alpha_{jj'}.
\]
We refer to this approach as GEE-UN.

With each of the five forms of working correlation matrices, we could
solve~(\ref{eq1-gee1}) using the Newton--Raphson method to obtain the
corresponding parameter estimates $\widehat{\bolds{\beta}}$. With the
exchangeable or unstructured
correlation structure, we iteratively update $\alpha$ from its moment
estimator and $\bolds{\beta}$ from equation (\ref{eq1-gee1})
[\citet{LiaZeg86}]. According\vspace*{1pt} to the standard theory of GEE,
the variance of the estimated $\widehat{\bolds{\beta}}$ has
the usual sandwich form $A_{N}^{-1}B_{N}A_{N}^{-1}$, where
\begin{eqnarray*}
A_{N} & = & \sum_{i=1}^{N}D_{i}'V_{i}D_{i},
\\
B_{N} & = & \sum_{i=1}^{N}D_{i}'V_{i}^{-1}
\bigl(\mathbf{Y}_{i}-\bolds{\mu}_{i}^{M}
\bigr) \bigl(\mathbf{Y}_{i}-\bolds{\mu}_{i}^{M}
\bigr)'V_{i}^{-1}D_{i}.
\end{eqnarray*}

\subsection{Marginal covariate effect}\label{sec2.3}

We note that the regression parameters in the GLMM and GEE are not
directly comparable as they have different interpretations. The former
is interpreted as the ``subject-specific effect'' conditional
on a subject~$i$, while the latter is the ``population-averaged
effect'' or ``marginal effect'' [\citet{ZegLiaAlb88}]. Thus, GLMM and GEE are not compatible for
nonidentity link functions. In other words, if the GLMM is true,
the marginal expectation by integrating out the random effects $\mathbf{b}_{i}$
may not preserve the linear additive form of the covariates. However,
for binary regression with a probit link and random intercept, GLMM
and GEE are compatible. We adopt a probit random effects model for
both the simulations and example analysis.

Let $\Phi$ and $\phi$ be the c.d.f. and p.d.f. of the standard normal
distribution. Consider the generalized linear mixed effects model
with a probit link and a random intercept only,
\begin{eqnarray*}
\Pr(Y_{ij}=1\mid\mathbf{X}_{ij},b_{i}) & = &
\Phi\bigl(\mathbf{X}_{ij}'\bolds{\gamma}+b_{i}\bigr),
\\
b_{i} & \sim& N\bigl(0,\sigma_{b}^{2}\bigr).
\end{eqnarray*}
By integrating out $b_{i}$, the marginal probability of $Y_{ij}$
is computed as follows:
\begin{eqnarray*}
\Pr(Y_{ij}=1\mid\mathbf{X}_{ij}) & = & \int
_{-\infty}^{+\infty}\Pr(Y_{ij}=1\mid
\mathbf{X}_{ij},b_{i})f(b_{i})\,db_{i}
\\
& = & \int_{-\infty}^{+\infty}\Phi\bigl(
\mathbf{X}_{ij}'\bolds{\gamma}+b_{i}
\bigr)\frac{1}{\sigma_{b}}\phi\biggl(\frac{b_{i}}{\sigma_{b}} \biggr)\,db_{i}
\\
& = & \Phi\biggl(\frac{\mathbf{X}_{ij}'\bolds{\gamma}}{\sqrt{1+\sigma
_{b}^{2}}} \biggr).
\end{eqnarray*}
While GLMM estimates $\Pr(Y_{ij}=1\mid\mathbf{X}_{ij},b_{i})$,
GEE estimates $\Pr(Y_{ij}=1\mid \mathbf{X}_{ij})$.
The latter is a probit regression model as well, with the regression
coefficients, $\bolds{\gamma}/\sqrt{1+\sigma_{b}^{2}}$. This
allows us to compare the performance of GLMM and GEE by comparing
the marginal effects of the covariates, which is our interest in the
dating violence analysis of the NEXT study.

\section{Model identifiability and test for zero inflation}\label{sec3}

In general, zero-\break inflated models are mixtures of two parametric parts,
a point mass at zero (equivalently, a~binary distribution with $p=0$)
and a parametric distribution for the nonstructural zero part. Typically,
zero-inflated models are identified by observing a larger number of
zeros than would be consistent with the parametric model. For example,
with Poisson or binomial outcomes, one can observe excessive proportion
of zeros with a histogram. For a single binary outcome, zero inflation
cannot be distinguished from rare events, unless covariate dependence
is introduced. When there is a continuous covariate $X$, zero inflation
is identified because of the linear effect of $X$ on the binary response
through a known link function. \citet{FolLam91} proved
a weaker sufficient condition for identifiability when covariates
are all categorical: to identify a two component mixture of logistic
regressions with a binary response, the covariate vector needs to
take at least 7 distinct values. \citet{KelAnd08} also used
the same argument to prove the identifiability of zero-inflated ordinal
regression. Single\vspace*{1pt} binary outcome can be seen as a special case of
our proposed model with $J=1$ and $\sigma_{b}^{2}=0$. As more information
is available with $J>1$, our model is also identified under Follmann
and Lambert's condition.

\citet{DioDioDup11} proved the model identifiability for the
zero-inflated binary regression with at least one continuous covariate.
Using a similar technique, we can prove our model identifiability.
For GEE with a probit link, consider $(\bolds{\beta}',p)$
and $(\bolds{\beta}^{*\prime},p^{*})$ to be two parameter vectors
that yield the same conditional mean $E(Y_{ij}\mid \mathbf{X}_{ij})$,
that is,
%
\begin{equation}
p\Phi\bigl(\mathbf{X}'\bolds{\beta} \bigr)=p^{*}
\Phi\bigl(\mathbf{X}'\bolds{\beta}^{*}
\bigr).\label{eq1-iden}
\end{equation}
Equivalently, $\frac{p}{p^{*}}=\frac{\Phi(\mathbf{X}'\bolds{\beta}^{*}
)}{\Phi(\mathbf{X}'\bolds{\beta} )}$.
Suppose the $l$th component of $\mathbf{X}$ (i.e., $x_{l}$)
is continuous, then we can take the partial derivative with respect to
$x_{l}$, which yields
%
%
\begin{eqnarray}\label{eq2-iden}
&& 0  =  \frac{\phi(\mathbf{X}'\bolds{\beta}^{*} )\beta_{l}^{*}\Phi
(\mathbf{X}'\bolds{\beta} )-\phi(\mathbf{X}'\bolds{\beta} )\beta
_{l}\Phi(\mathbf{X}'\bolds{\beta}^{*} )}{\Phi^{2} (\mathbf{X}'\bolds
{\beta} )}\nonumber
\\
&&\quad \Longleftrightarrow\quad \frac{\beta_{l}^{*}}{\beta_{l}} =  \frac{\Phi
(\mathbf{X}'\bolds{\beta}^{*} )\phi(\mathbf{X}\bolds{\beta} )}{\Phi
(\mathbf{X}'\bolds{\beta} )\phi(\mathbf{X}'\bolds{\beta}^{*} )}
\\
&&\quad \Longleftrightarrow\quad\frac{\beta_{l}^{*}}{\beta_{l}}  =  \frac{p\phi
(\mathbf{X}'\bolds{\beta} )}{p^{*}\phi(\mathbf{X}'\bolds{\beta}^{*})}.\nonumber
\end{eqnarray}
Taking the partial derivative on both sides of (\ref{eq2-iden}) with
respect to $x_{l}$, and with some algebra, it follows that $\mathbf
{X}'\bolds{\beta}=\mathbf{X}'\bolds{\beta}^{*}$,
and hence $\bolds{\beta}=\bolds{\beta}^{*}$. From (\ref{eq1-iden}),
we further get $p=p^{*}$. This proves the identifiability GEE-CI
and GEE-MI. In GEE-UN, the association parameters
are indeed obtained from a moment estimator of the correlation between $Y_{ij}$
and $Y_{ij'}$. Since the mean model is identified, the variance
and correlation are also identified. For exchangeable working correlation,
the association parameter is the ``average'' correlation
between all $Y_{ij}$ and $Y_{ij'}$ pairs with $j\neq j'$,
which is identifiable as well.

We now prove the identifiability of the random effects model (\ref{eq0-glmm})
with a probit link. With normally distributed random effects, the
mean of $Y_{ij}$ could be marginalized as $\Phi(\frac{\mathbf{X}'\bolds
{\gamma}}{\sqrt{1+\mathbf{Z}'\Delta\mathbf{Z}}} )$,
where $\Delta$ is the variance--covariance matrix of random effects~$\mathbf{b}_{i}$. We further assume that a continuous covariate
is contained in $\mathbf{X}$ but not in~$\mathbf{Z}$. Then
the same argument of (\ref{eq2-iden}) still applies by denoting $\frac
{\bolds{\gamma}}{\sqrt{1+\mathbf{Z}'\Delta\mathbf{Z}}}$
as $\bolds{\beta}$, which proves the identifiability of the
regression coefficients up to a scale. Now it suffices to prove the
identifiability of $\Delta$. Denote $\alpha_{jj'}$ as the
correlation coefficient of $Y_{ij}$ and $Y_{ij'}$ ($j\neq j'$)
given $Z=1$. Note that from Section~\ref{sec2.2}, we have
\[
\cov(Y_{ij},Y_{ij'})=\alpha_{jj'}p\sqrt{
\mu_{ij}^{Z}\bigl(1-\mu_{ij}^{Z}\bigr)
\mu_{ij'}^{Z}\bigl(1-\mu_{ij'}^{Z}\bigr)}+
\mu_{ij}^{Z}\mu_{ij'}^{Z}p(1-p).
\]
Since $\bolds{\beta}$ is identifiable, $\mu_{ij}^{Z}=\Phi(\mathbf
{X}_{ij}'\bolds{\beta})$
is also identifiable. Therefore, if two parameter vectors $\bolds{\theta
}=(\bolds{\gamma}',p,\Delta)'$
and $\bolds{\theta}^{*}=(\bolds{\gamma}^{*\prime},p^{*},\Delta^{*})'$
lead to the same $\cov(Y_{ij},Y_{ij'})$ and $EY_{ij}$, $\alpha_{jj'}$
must be the same. Furthermore, the regular GLMM is identifiable, suggesting
that the correlation structure $\alpha_{jj'}$ conditional
on $Z=1$ is uniquely defined by $\Delta$. Hence, we prove $\Delta=\Delta^{*}$,
and, consequently, the identifiability of the ML estimator is established.

We also note that when $\sigma_{b}^{2}=0$ and no covariates are available,
the repeated binary counts could be collapsed into a binomial distribution.
The problem then reduces to the zero-inflated binomial model, which
is clearly identifiable. In the presence of the random effects, collapsing
the binary counts leads to an over-dispersed binomial distribution.
\citet{HalBer02} discussed the zero-inflated beta-binomial
model, where the over-dispersion is controlled by a beta distributed
random intercept. Our model assumes that the over-dispersion comes
from a normal distributed random intercept.

Another way to view the proposed model is a mixture of random effect
distributions. Recall that $Y_{ij}$ follows a Bernoulli distribution
with probability $\pi_{ij}$, given by
\[
g \bigl\{ \pi_{ij}(\mathbf{b}_{i}) \bigr\} =
\mathbf{X}_{ij}'\bolds{\gamma}+
\mathbf{Z}_{ij}'\mathbf{b}_{i}.
\]
Instead of introducing the latent class $Z_{i}$, we assume that $\mathbf{b}_{i}$
is a mixture of normal distribution and a point mass at $-\infty$:
\[
\mathbf{b}_{i}=\cases{ \mathrm{MVN}(0,\Delta), &\quad with probability
$p$,
\cr
-\infty, &\quad with probability $1-p$.}
\]
When $\mathbf{b}_{i}=-\infty$, the probability $\pi_{ij}$ is
always 0 for $j=1,\ldots,J$, so $\mathbf{Y}_{i}$ is the structural
zero. It is easy to show that the likelihood is exactly the same as
the proposed model.

In practice, one may wish to test for the existence of zero inflation,
which can be performed under the parametric model framework. The likelihood
ratio statistic is given by
\[
\Lambda=2(l_{1}-l_{0}),
\]
where $l_{1}$ is the maximized log-likelihood for the zero-inflated
model, and $l_{0}$ is the maximized log-likelihood for the ordinary
GLMM. As the null hypothesis ($p=0$) is on the boundary of the parameter
space, the asymptotic null distribution of $\Lambda$ is a mixture
of $\chi_{1}^{2}$ and point mass at 0, with equal mixture probabilities
[\citet{SelLia87}]. Theoretically, we could also construct a score
test statistic similar to the test proposed by \citet{van95}
for zero inflation in a Poisson distribution. However, for our problem,
the likelihood function involves intractable integrals, making the
score and information matrix both difficult to evaluate. So in our
application, we apply the likelihood ratio test.

\section{Simulation studies}\label{sec4}

Motivated by the NEXT study, the data generation for the simulation
studies mimics the real example. To evaluate the statistical properties
of the above methods, simulation studies of the true model and a misspecified
model were run with two different levels of within-cluster correlation.
A sample size of $N=2000$ with a cluster size of $J=5$ questions
is considered. The simulations were repeated 5000 times to compare
the performance of the naive estimator (GLMM, where the zero-inflation
is ignored), the maximum-likelihood (ML) estimator and the five GEE estimators
(GEE-MI, GEE-CI, GEE-ME, GEE-CE and GEE-UN). We calculated the average
(Mean) and standard deviation (SD) of the estimated parameters, average
of the estimated standard errors (SE) and 95\% CI coverage rates
(COVER) based on the Wald intervals to evaluate the robustness and
efficiency of the GEE and the
maximum-likelihood approaches. Twenty Gaussian--Hermite quadrature
points were used for computing the GLMM and ML estimators. We also
tried 10 and 40 quadrature
points as well as the adaptive quadrature with 250 simulated data sets.
In our simulations,
the results are very similar for differing number of quadrature points.
Our experience for generalized
linear mixed models with the logit link function is that Gaussian quadrature
works very well, and in most situations AGQ is not needed. In terms of
numerical efficiency,
we found that the computation time for AGQ is about 10--20 times longer
than the fixed quadrature.

%
\begin{table}[b]
\tabcolsep=0pt
\caption{The mean of 5000 simulations of estimated coefficients (Mean),
empirical standard deviation (SD), average standard error (SE) and the 95\%
interval coverage rate (COVER) for the maximum-likelihood, naive and GEE
methods of the correctly specified model with $\sigma_{b}=0.5$, $N=2000$}\label{tab1}
\begin{tabular*}{\tablewidth}{@{\extracolsep{\fill}}@{}lcd{2.3}d{2.3}ccc@{}}
\hline
& \textbf{Parameter\tabnoteref{tt1}} & \multicolumn{1}{c}{\textbf{True}} & \multicolumn{1}{c}{\textbf{Mean}}
& \multicolumn{1}{c}{\textbf{SD}} & \textbf{SE} & \textbf{COVER}
\\
\hline
ML & $\operatorname{Pr}(Z=1)$ & 0.700 & 0.700 & 0.014 & 0.014 & 0.949 \\
& $\sigma_{b}$ & 0.500 & 0.499 & 0.040 & 0.039 & 0.953 \\
& $\beta_{0}$ & 0.000 & 0.000 & 0.039 & 0.040 & 0.959 \\
& $\beta_{1}$ & 0.894 & 0.895 & 0.027 & 0.027 & 0.949 \\
& $\beta_{2}$ & -0.447 & -0.448 & 0.051 & 0.051 & 0.951 \\
& $\beta_{3}$ & -0.358 & -0.358 & 0.050 & 0.051 & 0.956 \\
& $\beta_{4}$ & 0.179 & 0.179 & 0.051 & 0.050 & 0.949 \\
& $\beta_{5}$ & 0.358 & 0.358 & 0.050 & 0.051 & 0.948
\\[3pt]
GLMM & $\sigma_{b}$ & 0.500 & 1.352 & 0.047 & 0.046 & 0.000\\
& $\beta_{0}$ & 0.000 & -0.444 & 0.029 & 0.030 & 0.000 \\
& $\beta_{1}$ & 0.894 & 0.570 & 0.028 & 0.025 & 0.000 \\
& $\beta_{2}$ & -0.447 & -0.301 & 0.034 & 0.034 & 0.013 \\
& $\beta_{3}$ & -0.358 & -0.239 & 0.033 & 0.034 & 0.060 \\
& $\beta_{4}$ & 0.179 & 0.114 & 0.032 & 0.032 & 0.489 \\
& $\beta_{5}$ & 0.358 & 0.224 & 0.031 & 0.032 & 0.015
\\
\hline
\end{tabular*}
\end{table}

The estimated
parameters for ML and GLMM methods were marginalized, as we described
in Section~\ref{sec2.3}. In the following sections, we evaluate the performance
of the maximum likelihood and GEE under a correctly specified and
a misspecified model. Additional simulation results are reported
in the supplementary material [\citet{Fuletal14}], including (a) the
performance of the
proposed model with a smaller sample size ($N=500$); (b) sensitivity
of assuming a constant zero-inflation probability when the probability
is affected by covariates; (c) performance of zero-inflated beta-binomial
model.

\subsection{Simulation one: Correctly specified model}\label{sec4.1}

We generated a continuous subject-level covariate $X_{i}$ from a
standard normal distribution and categorical covariate $Q_{ij}=
1,\ldots,5$ to denote the questions for each subject. The zero-inflation
indicator $Z_{i}$ was generated from $\operatorname{Bernoulli}(p)$ with $p=0.7$.
The outcome $Y_{ij}$ was generated from a probit random effects model:
\begin{eqnarray*}
&& P(Y_{ij}=1\mid X_{i},Q_{ij},b_{i},Z_{i}=1)
\\
&&\qquad  =  \Phi\bigl\{ \gamma_{0}+\gamma_{1}X_{i}+
\gamma_{2}I(Q_{ij}=2)+\gamma_{3}I(Q_{ij}=3)
\\
&&\hspace*{70pt}{}  +\gamma_{4}I(Q_{ij}=4)+\gamma_{5}I(Q_{ij}=5)+b_{i}
\bigr\} ,
\end{eqnarray*}
where $I(\cdot)$ is the indicator function and $b_{i}$ is the random
intercept following a normal distribution $N(0,\sigma_{b}^{2})$.
We fixed the regression parameters $\bolds{\gamma}= (\gamma_{0},\gamma
_{1},\gamma_{2},\gamma_{3},\gamma_{4},\gamma_{5} )'=(0,1,-0.5,-0.4,0.2,0.4)'$.
The variance component $\sigma_{b}^{2}$ was taken to be $0.5^{2}$
and $1.5^{2}$, respectively, to reflect weak (Pearson correlation of
about 0.1) and strong (Pearson correlation of about 0.45) within-cluster
correlations. The simulation results are shown in Tables~\ref{tab1}
and \ref{tab2}, where the true regression parameters are the marginal
covariate effects given by $\bolds{\beta}=\frac{\bolds{\gamma}}{\sqrt
{1+\sigma_{b}^{2}}}$.

\setcounter{table}{0}
\begin{table}
\tabcolsep=0pt
\caption{(Continued)}
\begin{tabular*}{\tablewidth}{@{\extracolsep{\fill}}@{}lcd{2.3}d{2.3}ccc@{}}
\hline
& \textbf{Parameter\tabnoteref{tt1}} & \multicolumn{1}{c}{\textbf{True}} & \multicolumn{1}{c}{\textbf{Mean}}
& \multicolumn{1}{c}{\textbf{SD}} & \textbf{SE} & \textbf{COVER}
\\
\hline
GEE-MI & $\operatorname{Pr}(Z=1)$ & 0.700 & 0.704 & 0.041 & 0.040 & 0.951 \\
& $\beta_{0}$ & 0.000 & -0.002 & 0.096 & 0.096 & 0.951 \\
& $\beta_{1}$ & 0.894 & 0.897 & 0.061 & 0.061 & 0.947 \\
& $\beta_{2}$ & -0.447 & -0.448 & 0.060 & 0.060 & 0.950 \\
& $\beta_{3}$ & -0.358 & -0.358 & 0.056 & 0.057 & 0.952 \\
& $\beta_{4}$ & 0.179 & 0.179 & 0.055 & 0.054 & 0.952 \\
& $\beta_{5}$ & 0.358 & 0.358 & 0.059 & 0.059 & 0.946
\\[3pt]
GEE-CI & $\operatorname{Pr}(Z=1)$ & 0.700 & 0.701 & 0.029 & 0.029 & 0.948 \\
& $\beta_{0}$ & 0.000 & 0.001 & 0.069 & 0.070 & 0.955 \\
& $\beta_{1}$ & 0.894 & 0.897 & 0.044 & 0.043 & 0.953 \\
& $\beta_{2}$ & -0.447 & -0.449 & 0.055 & 0.056 & 0.953 \\
& $\beta_{3}$ & -0.358 & -0.359 & 0.053 & 0.054 & 0.958 \\
& $\beta_{4}$ & 0.179 & 0.179 & 0.052 & 0.052 & 0.951 \\
& $\beta_{5}$ & 0.358 & 0.360 & 0.056 & 0.056 & 0.946
\\[3pt]
GEE-ME & $\operatorname{Pr}(Z=1)$ & 0.700 & 0.701 & 0.032 & 0.032 & 0.953 \\
& $\beta_{0}$ & 0.000 & 0.001 & 0.077 & 0.078 & 0.953 \\
& $\beta_{1}$ & 0.894 & 0.898 & 0.049 & 0.049 & 0.949 \\
& $\beta_{2}$ & -0.447 & -0.449 & 0.058 & 0.058 & 0.955 \\
& $\beta_{3}$ & -0.358 & -0.359 & 0.055 & 0.056 & 0.956 \\
& $\beta_{4}$ & 0.179 & 0.180 & 0.054 & 0.054 & 0.952 \\
& $\beta_{5}$ & 0.358 & 0.359 & 0.057 & 0.058 & 0.950
\\[3pt]
GEE-CE & $\operatorname{Pr}(Z=1)$ & 0.700 & 0.701 & 0.029 & 0.029 & 0.950 \\
& $\beta_{0}$ & 0.000 & 0.001 & 0.069 & 0.070 & 0.955 \\
& $\beta_{1}$ & 0.894 & 0.897 & 0.044 & 0.043 & 0.953 \\
& $\beta_{2}$ & -0.447 & -0.449 & 0.055 & 0.056 & 0.954 \\
& $\beta_{3}$ & -0.358 & -0.359 & 0.052 & 0.054 & 0.958 \\
& $\beta_{4}$ & 0.179 & 0.179 & 0.052 & 0.052 & 0.951 \\
& $\beta_{5}$ & 0.358 & 0.360 & 0.056 & 0.056 & 0.946
\\[3pt]
GEE-UN & $\operatorname{Pr}(Z=1)$ & 0.700 & 0.702 & 0.036 & 0.036 & 0.952 \\
& $\beta_{0}$ & 0.000 & 0.000 & 0.085 & 0.085 & 0.952 \\
& $\beta_{1}$ & 0.894 & 0.897 & 0.053 & 0.053 & 0.948 \\
& $\beta_{2}$ & -0.447 & -0.448 & 0.058 & 0.059 & 0.954 \\
& $\beta_{3}$ & -0.358 & -0.358 & 0.055 & 0.056 & 0.954 \\
& $\beta_{4}$ & 0.179 & 0.179 & 0.055 & 0.054 & 0.952 \\
& $\beta_{5}$ & 0.358 & 0.359 & 0.058 & 0.058 & 0.948 \\
\hline
\end{tabular*}
\tabnotetext{tt1}{$P(Y_{ij}=1\mid X_{i},Q_{ij},Z_{i}=1)=\Phi \{ \beta
_{0}+\beta_{1}X_{ij}+\beta_{2}I(Q_{ij}=2)+
\beta_{3}I(Q_{ij}=3)+\beta_{4}I(Q_{ij}=4)+\beta_{5}I(Q_{ij}=5) \}$.}\vspace*{-3pt}
\end{table}

%
\begin{table}
\tabcolsep=0pt
\caption{The mean of 5000 simulations of estimated coefficients (Mean),
empirical standard deviation (SD), average standard error (SE) and the
95\% interval coverage rate (COVER) for the maximum-likelihood, naive and GEE
methods of the correctly specified model with $\protect\sigma_{b}=1.5$, $N=2000$}\label{tab2}
\begin{tabular*}{\tablewidth}{@{\extracolsep{\fill}}@{}lcd{2.3}d{2.3}ccc@{}}
\hline
& \textbf{Parameter\tabnoteref{tt2}} & \multicolumn{1}{c}{\textbf{True}} & \multicolumn{1}{c}{\textbf{Mean}}
& \multicolumn{1}{c}{\textbf{SD}} & \textbf{SE} & \textbf{COVER}
\\
\hline
ML & $\operatorname{Pr}(Z=1)$ & 0.700 & 0.701 & 0.024 & 0.024 & 0.951 \\
& $\sigma_{b}$ & 1.500 & 1.503 & 0.088 & 0.089 & 0.955 \\
& $\beta_{0}$ & 0.000 & -0.001 & 0.053 & 0.053 & 0.950 \\
& $\beta_{1}$ & 0.555 & 0.555 & 0.033 & 0.033 & 0.949 \\
& $\beta_{2}$ & -0.277 & -0.278 & 0.037 & 0.038 & 0.952 \\
& $\beta_{3}$ & -0.222 & -0.222 & 0.037 & 0.037 & 0.949 \\
& $\beta_{4}$ & 0.111 & 0.111 & 0.036 & 0.036 & 0.954 \\
& $\beta_{5}$ & 0.222 & 0.222 & 0.037 & 0.037 & 0.951
\\[3pt]
GLMM & $\sigma_{b}$ & 1.500 & 2.248 & 0.071 & 0.073 & 0.000\\
& $\beta_{0}$ & 0.000 & -0.419 & 0.032 & 0.029 & 0.000 \\
& $\beta_{1}$ & 0.555 & 0.384 & 0.029 & 0.026 & 0.000 \\
& $\beta_{2}$ & -0.277 & -0.204 & 0.027 & 0.027 & 0.230 \\
& $\beta_{3}$ & -0.222 & -0.163 & 0.027 & 0.027 & 0.387 \\
& $\beta_{4}$ & 0.111 & 0.079 & 0.025 & 0.026 & 0.763 \\
& $\beta_{5}$ & 0.222 & 0.155 & 0.025 & 0.026 & 0.271
\\[3pt]
GEE-MI & $\operatorname{Pr}(Z=1)$ & 0.700 & 0.712 & 0.083 & 0.088 & 0.949 \\
& $\beta_{0}$ & 0.000 & -0.002 & 0.161 & 0.170 & 0.966 \\
& $\beta_{1}$ & 0.555 & 0.560 & 0.072 & 0.075 & 0.962 \\
& $\beta_{2}$ & -0.277 & -0.279 & 0.050 & 0.050 & 0.949 \\
& $\beta_{3}$ & -0.222 & -0.223 & 0.046 & 0.046 & 0.950 \\
& $\beta_{4}$ & 0.111 & 0.111 & 0.040 & 0.040 & 0.954 \\
& $\beta_{5}$ & 0.222 & 0.223 & 0.047 & 0.049 & 0.953
\\[3pt]
GEE-CI & $\operatorname{Pr}(Z=1)$ & 0.700 & 0.709 & 0.064 & 0.064 & 0.954 \\
& $\beta_{0}$ & 0.000 & -0.006 & 0.121 & 0.123 & 0.969 \\
& $\beta_{1}$ & 0.555 & 0.556 & 0.056 & 0.057 & 0.958 \\
& $\beta_{2}$ & -0.277 & -0.278 & 0.045 & 0.045 & 0.951 \\
& $\beta_{3}$ & -0.222 & -0.222 & 0.042 & 0.042 & 0.946 \\
& $\beta_{4}$ & 0.111 & 0.111 & 0.038 & 0.039 & 0.952 \\
& $\beta_{5}$ & 0.222 & 0.223 & 0.044 & 0.045 & 0.951
\\[3pt]
GEE-ME & $\operatorname{Pr}(Z=1)$ & 0.700 & 0.706 & 0.058 & 0.057 & 0.947 \\
& $\beta_{0}$ & 0.000 & -0.001 & 0.115 & 0.114 & 0.954 \\
& $\beta_{1}$ & 0.555 & 0.557 & 0.053 & 0.053 & 0.952 \\
& $\beta_{2}$ & -0.277 & -0.279 & 0.045 & 0.044 & 0.948 \\
& $\beta_{3}$ & -0.222 & -0.223 & 0.042 & 0.042 & 0.947 \\
& $\beta_{4}$ & 0.111 & 0.111 & 0.039 & 0.039 & 0.955 \\
& $\beta_{5}$ & 0.222 & 0.223 & 0.043 & 0.044 & 0.953 \\
\hline
\end{tabular*}
\end{table}

\setcounter{table}{1}
\begin{table}
\tabcolsep=0pt
\caption{(Continued)}
\begin{tabular*}{\tablewidth}{@{\extracolsep{\fill}}@{}lcd{2.3}d{2.3}ccc@{}}
\hline
& \textbf{Parameter\tabnoteref{tt2}} & \multicolumn{1}{c}{\textbf{True}} & \multicolumn{1}{c}{\textbf{Mean}}
& \multicolumn{1}{c}{\textbf{SD}} & \textbf{SE} & \textbf{COVER}
\\
\hline
GEE-CE & $\operatorname{Pr}(Z=1)$ & 0.700 & 0.707 & 0.056 & 0.056 & 0.951 \\
& $\beta_{0}$ & 0.000 & -0.003 & 0.111 & 0.111 & 0.956 \\
& $\beta_{1}$ & 0.555 & 0.556 & 0.052 & 0.051 & 0.951 \\
& $\beta_{2}$ & -0.277 & -0.278 & 0.043 & 0.043 & 0.952 \\
& $\beta_{3}$ & -0.222 & -0.222 & 0.041 & 0.041 & 0.950 \\
& $\beta_{4}$ & 0.111 & 0.111 & 0.038 & 0.038 & 0.955 \\
& $\beta_{5}$ & 0.222 & 0.223 & 0.043 & 0.043 & 0.954
\\[3pt]
GEE-UN & $\operatorname{Pr}(Z=1)$ & 0.700 & 0.709 & 0.068 & 0.068 & 0.950 \\
& $\beta_{0}$ & 0.000 & -0.003 & 0.132 & 0.134 & 0.959 \\
& $\beta_{1}$ & 0.555 & 0.558 & 0.060 & 0.060 & 0.952 \\
& $\beta_{2}$ & -0.277 & -0.278 & 0.046 & 0.046 & 0.948 \\
& $\beta_{3}$ & -0.222 & -0.223 & 0.044 & 0.043 & 0.948 \\
& $\beta_{4}$ & 0.111 & 0.111 & 0.039 & 0.039 & 0.951 \\
& $\beta_{5}$ & 0.222 & 0.223 & 0.045 & 0.046 & 0.952 \\
\hline
\end{tabular*}
\tabnotetext{tt2}{$P(Y_{ij}=1\rrvert X_{i},Q_{ij},Z_{i}=1)=\Phi \{
\beta_{0}+\beta_{1}X_{ij}+\beta_{2}I(Q_{ij}=2)
+ \beta_{3}I(Q_{ij}=3)+\beta_{4}I(Q_{ij}=4)+\beta_{5}I(Q_{ij}=5) \}$.}
\end{table}

Both the ML and the five GEE methods perform reasonably well, in
terms of small bias and good CI coverage rate. GLMM is seriously biased
with poor CI\vadjust{\goodbreak} coverage. It can be seen that the ML method is the most
efficient, as it makes use of the full distributional assumption of
the observed data. On the contrary, GEE only relies on the first moments
of the outcome. In estimating $p$, the zero-inflated probability,
the SEs of the GEE approaches are more than twice as large as the
SE of the ML method. The SEs for other parameters are also significantly
smaller for the ML method.

Of the five GEE methods, we found that GEE-CE is the most efficient
with the smallest SE, while GEE-MI is the least efficient. By exploiting
the correlation structure induced by the zero-inflation process, the
conditional independence and exchangeable working correlation both
gain a substantial amount of efficiency, compared to their marginal
counterparts. This result is consistent with the simulation results in
\citet{HalZha04}.
The SEs for GEE-CE and GEE-CI are quite close, implying
that adding working dependence to the outcome given $Z_{i}=1$ would
not help much as long as the dependence due to zero inflation is accounted
for. We did observe a bigger improvement of GEE-CE versus GEE-CI for
the strong correlation case.
But the improvement of GEE-CI versus GEE-MI is even larger. Therefore,
we recommend
that it is more important to make use of the zero-inflation structure
in the GEE estimators.
Although GEE-UN has the most flexible form of working correlation,
it is not as efficient as GEE-CI or GEE-CE, probably due to estimating
a larger number of nuisance parameters. We found that GEEs may occasionally
not have a solution or have a boundary solution ($\hat{p}=1$) in
about 1--2\% of the simulations with $\sigma_{b}=1.5$. Our experience
is that nonconvergence or boundary solutions occur more often when
the covariate effects are weaker, the within-cluster correlation is
stronger, the true zero-inflation probability is closer to~1, or the
model is more severely misspecified.\

\subsection{Simulation two: Model misspecification}\label{sec4.2}

We consider a misspecified model where only the first three questions
are correlated and the last two are independent. The data generation
for $Y_{ij}$ with $j=1,2,3$ was the same as in Section~3.1, but
$Y_{ij}$ for $j=4,5$ was generated as follows:
\begin{eqnarray*}
&& P(Y_{ij}=1\mid X_{i},Q_{ij},b_{i},Z_{i}=1)
\\
&&\qquad =
\Phi\biggl\{ \frac{\gamma_{0}+\gamma_{1}X_{ij}+\gamma
_{4}I(Q_{ij}=4)+\gamma_{5}I(Q_{ij}=5)}{\sqrt{1+\sigma_{b}^{2}}} \biggr
\} .
\end{eqnarray*}
The random intercept $b_{i}$ was not added to the last two questions,
but a factor of $\sqrt{1+\sigma_{b}^{2}}$ was divided to the coefficients
to keep the marginalized regression coefficients the same. In this
case, the ML estimator is from a misspecified model since the random
intercept model imposes correlation among all the questions. For GEE,
only the working correlation is misspecified, while the first moment
of $Y_{ij}$ is still correct. The simulation results are presented
in Tables~\ref{tab3} and \ref{tab4}.

%
\begin{table}[b]
\tabcolsep=0pt
\caption{The mean of 5000 simulations of estimated coefficients (Mean),
empirical standard deviation (SD), average standard error (SE) and the
95\%
interval coverage rate (COVER) for the maximum-likelihood, naive and GEE
methods of the misspecified model with $\protect\sigma_{b}=0.5$, $N=2000$}\label{tab3}
\begin{tabular*}{\tablewidth}{@{\extracolsep{\fill}}@{}lcd{2.3}d{2.3}ccc@{}}
\hline
& \textbf{Parameter\tabnoteref{tt3}} & \multicolumn{1}{c}{\textbf{True}} & \multicolumn{1}{c}{\textbf{Mean}}
& \multicolumn{1}{c}{\textbf{SD}} & \textbf{SE} & \textbf{COVER}
\\
\hline
ML & $\operatorname{Pr}(Z=1)$ & 0.700 & 0.706 & 0.013 & 0.013 & 0.925\\
& $\sigma_{b}$ & 0.500 & 0.273 & 0.047 & 0.047 & 0.000\\
& $\beta_{0}$ & 0.000 & -0.008 & 0.038 & 0.039 & 0.953\\
& $\beta_{1}$ & 0.894 & 0.899 & 0.024 & 0.024 & 0.948\\
& $\beta_{2}$ & -0.447 & -0.446 & 0.051 & 0.053 & 0.963\\
& $\beta_{3}$ & -0.358 & -0.357 & 0.049 & 0.053 & 0.964\\
& $\beta_{4}$ & 0.179 & 0.177 & 0.053 & 0.052 & 0.949\\
& $\beta_{5}$ & 0.358 & 0.355 & 0.053 & 0.053 & 0.949
\\[3pt]
GLMM & $\sigma_{b}$ & 0.500 & 1.176 & 0.042 & 0.040 &0.000\\
& $\beta_{0}$ & 0.000 & -0.441 & 0.029 & 0.030 & 0.000\\
& $\beta_{1}$ & 0.894 & 0.570 & 0.028 & 0.024 & 0.000\\
& $\beta_{2}$ & -0.447 & -0.303 & 0.035 & 0.036 & 0.019\\
& $\beta_{3}$ & -0.358 & -0.241 & 0.034 & 0.035 & 0.080\\
& $\beta_{4}$ & 0.179 & 0.114 & 0.034 & 0.034 & 0.519\\
& $\beta_{5}$ & 0.358 & 0.223 & 0.033 & 0.034 & 0.021
\\[3pt]
GEE-MI & $\operatorname{Pr}(Z=1)$ & 0.700 & 0.704 & 0.041 & 0.040 & 0.951\\
& $\beta_{0}$ & 0.000 & -0.002 & 0.095 & 0.095 & 0.951\\
& $\beta_{1}$ & 0.894 & 0.896 & 0.059 & 0.060 & 0.949\\
& $\beta_{2}$ & -0.447 & -0.447 & 0.059 & 0.060 & 0.953\\
& $\beta_{3}$ & -0.358 & -0.357 & 0.056 & 0.057 & 0.953\\
& $\beta_{4}$ & 0.179 & 0.178 & 0.057 & 0.057 & 0.950\\
& $\beta_{5}$ & 0.358 & 0.358 & 0.061 & 0.062 & 0.948
\\
\hline
\end{tabular*}
\end{table}

\setcounter{table}{2}
\begin{table}
\tabcolsep=0pt
\caption{(Continued)}
\begin{tabular*}{\tablewidth}{@{\extracolsep{\fill}}@{}lcd{2.3}d{2.3}ccc@{}}
\hline
& \textbf{Parameter\tabnoteref{tt3}} & \multicolumn{1}{c}{\textbf{True}} & \multicolumn{1}{c}{\textbf{Mean}}
& \multicolumn{1}{c}{\textbf{SD}} & \textbf{SE} & \textbf{COVER}
\\
\hline
GEE-CI & $\operatorname{Pr}(Z=1)$ & 0.700 & 0.701 & 0.030 & 0.030 & 0.950\\
& $\beta_{0}$ & 0.000 & 0.000 & 0.070 & 0.071 & 0.955\\
& $\beta_{1}$ & 0.894 & 0.897 & 0.044 & 0.044 & 0.951\\
& $\beta_{2}$ & -0.447 & -0.449 & 0.055 & 0.056 & 0.956\\
& $\beta_{3}$ & -0.358 & -0.358 & 0.053 & 0.054 & 0.957\\
& $\beta_{4}$ & 0.179 & 0.179 & 0.055 & 0.055 & 0.949\\
& $\beta_{5}$ & 0.358 & 0.360 & 0.059 & 0.059 & 0.951
\\[3pt]
GEE-ME & $\operatorname{Pr}(Z=1)$ & 0.700 & 0.702 & 0.034 & 0.033 & 0.951\\
& $\beta_{0}$ & 0.000 & 0.000 & 0.079 & 0.080 & 0.952\\
& $\beta_{1}$ & 0.894 & 0.897 & 0.049 & 0.050 & 0.949\\
& $\beta_{2}$ & -0.447 & -0.448 & 0.058 & 0.058 & 0.955\\
& $\beta_{3}$ & -0.358 & -0.358 & 0.055 & 0.056 & 0.954\\
& $\beta_{4}$ & 0.179 & 0.179 & 0.057 & 0.057 & 0.952\\
& $\beta_{5}$ & 0.358 & 0.359 & 0.061 & 0.061 & 0.951
\\[3pt]
GEE-CE & $\operatorname{Pr}(Z=1)$ & 0.700 & 0.701 & 0.030 & 0.030 & 0.951\\
& $\beta_{0}$ & 0.000 & 0.000 & 0.070 & 0.071 & 0.955\\
& $\beta_{1}$ & 0.894 & 0.897 & 0.044 & 0.044 & 0.950\\
& $\beta_{2}$ & -0.447 & -0.448 & 0.055 & 0.056 & 0.955\\
& $\beta_{3}$ & -0.358 & -0.358 & 0.053 & 0.054 & 0.956\\
& $\beta_{4}$ & 0.179 & 0.179 & 0.055 & 0.055 & 0.950\\
& $\beta_{5}$ & 0.358 & 0.359 & 0.059 & 0.059 & 0.952
\\[3pt]
GEE-UN & $\operatorname{Pr}(Z=1)$ & 0.700 & 0.703 & 0.037 & 0.036 & 0.952\\
& $\beta_{0}$ & 0.000 & -0.001 & 0.086 & 0.087 & 0.951\\
& $\beta_{1}$ & 0.894 & 0.897 & 0.053 & 0.054 & 0.948\\
& $\beta_{2}$ & -0.447 & -0.448 & 0.058 & 0.059 & 0.953\\
& $\beta_{3}$ & -0.358 & -0.358 & 0.055 & 0.056 & 0.954\\
& $\beta_{4}$ & 0.179 & 0.179 & 0.057 & 0.057 & 0.950\\
& $\beta_{5}$ & 0.358 & 0.359 & 0.061 & 0.062 & 0.950\\
\hline
\end{tabular*}
\tabnotetext{tt3}{$P(Y_{ij}=1\mid X_{i},Q_{ij},Z_{i}=1)=\Phi \{
\beta_{0}+\beta_{1}X_{ij}+\beta_{2}I(Q_{ij}=2)+
\beta_{3}I(Q_{ij}=3)+\beta_{4}I(Q_{ij}=4)+\beta_{5}I(Q_{ij}=5) \}$.}
\end{table}

%
\begin{table}
\tabcolsep=0pt
\caption{The mean of 5000 simulations of estimated coefficients (Mean),
empirical standard deviation (SD), average standard error (SE) and the
95\%
interval coverage rate (COVER) for the maximum-likelihood, naive and GEE
methods of the misspecified model with $\protect\sigma_{b}=1.5$, $N=2000$}\label{tab4}
\begin{tabular*}{\tablewidth}{@{\extracolsep{\fill}}@{}lcd{2.3}d{2.3}ccc@{}}
\hline
& \textbf{Parameter\tabnoteref{tt4}} & \multicolumn{1}{c}{\textbf{True}} & \multicolumn{1}{c}{\textbf{Mean}}
& \multicolumn{1}{c}{\textbf{SD}} & \textbf{SE} & \textbf{COVER}
\\
\hline
ML & $\operatorname{Pr}(Z=1)$ & 0.700 & 0.730 & 0.015 & 0.015 & 0.472\\
& $\sigma_{b}$ & 1.500 & 0.624 & 0.037 & 0.038 & 0.000\\
& $\beta_{0}$ & 0.000 & -0.046 & 0.039 & 0.039 & 0.776\\
& $\beta_{1}$ & 0.555 & 0.561 & 0.024 & 0.024 & 0.941\\
& $\beta_{2}$ & -0.277 & -0.273 & 0.036 & 0.046 & 0.986\\
& $\beta_{3}$ & -0.222 & -0.218 & 0.036 & 0.045 & 0.987\\
& $\beta_{4}$ & 0.111 & 0.103 & 0.047 & 0.044 & 0.933\\
& $\beta_{5}$ & 0.222 & 0.208 & 0.046 & 0.045 & 0.934
\\[3pt]
GLMM & $\sigma_{b}$ & 1.500 & 1.201 & 0.038 & 0.039 &0.000\\
& $\beta_{0}$ & 0.000 & -0.416 & 0.029 & 0.030 & 0.000\\
& $\beta_{1}$ & 0.555 & 0.378 & 0.024 & 0.022 & 0.000\\
& $\beta_{2}$ & -0.277 & -0.206 & 0.027 & 0.034 & 0.425\\
& $\beta_{3}$ & -0.222 & -0.164 & 0.027 & 0.034 & 0.613\\
& $\beta_{4}$ & 0.111 & 0.079 & 0.034 & 0.033 & 0.816\\
& $\beta_{5}$ & 0.222 & 0.153 & 0.033 & 0.032 & 0.432
\\[3pt]
GEE-MI & $\operatorname{Pr}(Z=1)$ & 0.700 & 0.712 & 0.080 & 0.082 & 0.948\\
& $\beta_{0}$ & 0.000 & -0.005 & 0.153 & 0.159 & 0.963\\
& $\beta_{1}$ & 0.555 & 0.558 & 0.066 & 0.067 & 0.954\\
& $\beta_{2}$ & -0.277 & -0.278 & 0.048 & 0.049 & 0.946\\
& $\beta_{3}$ & -0.222 & -0.222 & 0.045 & 0.045 & 0.949\\
& $\beta_{4}$ & 0.111 & 0.111 & 0.054 & 0.054 & 0.951\\
& $\beta_{5}$ & 0.222 & 0.223 & 0.058 & 0.06 & 0.954
\\[3pt]
GEE-CI & $\operatorname{Pr}(Z=1)$ & 0.700 & 0.707 & 0.066 & 0.066 & 0.944\\
& $\beta_{0}$ & 0.000 & -0.001 & 0.129 & 0.130 & 0.966\\
& $\beta_{1}$ & 0.555 & 0.558 & 0.058 & 0.058 & 0.960\\
& $\beta_{2}$ & -0.277 & -0.279 & 0.046 & 0.046 & 0.948\\
& $\beta_{3}$ & -0.222 & -0.223 & 0.043 & 0.043 & 0.950\\
& $\beta_{4}$ & 0.111 & 0.112 & 0.053 & 0.052 & 0.950\\
& $\beta_{5}$ & 0.222 & 0.224 & 0.057 & 0.058 & 0.951
\\[3pt]
GEE-ME & $\operatorname{Pr}(Z=1)$ & 0.700 & 0.709 & 0.069 & 0.068 & 0.948\\
& $\beta_{0}$ & 0.000 & -0.004 & 0.132 & 0.133 & 0.951\\
& $\beta_{1}$ & 0.555 & 0.557 & 0.060 & 0.060 & 0.952\\
& $\beta_{2}$ & -0.277 & -0.278 & 0.046 & 0.046 & 0.946\\
& $\beta_{3}$ & -0.222 & -0.222 & 0.043 & 0.043 & 0.946\\
& $\beta_{4}$ & 0.111 & 0.112 & 0.054 & 0.054 & 0.950\\
& $\beta_{5}$ & 0.222 & 0.224 & 0.058 & 0.059 & 0.953
\\
\hline
\end{tabular*}
\end{table}

\setcounter{table}{3}
\begin{table}
\tabcolsep=0pt
\caption{(Continued)}
\begin{tabular*}{\tablewidth}{@{\extracolsep{\fill}}@{}lcd{2.3}d{2.3}ccc@{}}
\hline
& \textbf{Parameter\tabnoteref{tt4}} & \multicolumn{1}{c}{\textbf{True}} & \multicolumn{1}{c}{\textbf{Mean}}
& \multicolumn{1}{c}{\textbf{SD}} & \textbf{SE} & \textbf{COVER}
\\
\hline
GEE-CE & $\operatorname{Pr}(Z=1)$ & 0.700 & 0.709 & 0.067 & 0.066 & 0.946\\
& $\beta_{0}$ & 0.000 & -0.004 & 0.129 & 0.130 & 0.959\\
& $\beta_{1}$ & 0.555 & 0.556 & 0.058 & 0.058 & 0.954\\
& $\beta_{2}$ & -0.277 & -0.279 & 0.045 & 0.046 & 0.948\\
& $\beta_{3}$ & -0.222 & -0.223 & 0.043 & 0.043 & 0.948\\
& $\beta_{4}$ & 0.111 & 0.111 & 0.053 & 0.052 & 0.948\\
& $\beta_{5}$ & 0.222 & 0.224 & 0.057 & 0.058 & 0.951
\\[3pt]
GEE-UN & $\operatorname{Pr}(Z=1)$ & 0.700 & 0.711 & 0.073 & 0.073 & 0.951\\
& $\beta_{0}$ & 0.000 & -0.005 & 0.140 & 0.142 & 0.954\\
& $\beta_{1}$ & 0.555 & 0.557 & 0.061 & 0.061 & 0.952\\
& $\beta_{2}$ & -0.277 & -0.278 & 0.047 & 0.047 & 0.947\\
& $\beta_{3}$ & -0.222 & -0.222 & 0.044 & 0.044 & 0.948\\
& $\beta_{4}$ & 0.111 & 0.111 & 0.053 & 0.053 & 0.950\\
& $\beta_{5}$ & 0.222 & 0.223 & 0.058 & 0.059 & 0.953\\
\hline
\end{tabular*}
\tabnotetext{tt4}{$P(Y_{ij}=1\mid X_{i},Q_{ij},Z_{i}=1)=\Phi \{ \beta
_{0}+\beta_{1}X_{ij}+\beta_{2}I(Q_{ij}=2)
+ \beta_{3}I(Q_{ij}=3)+\beta_{4}I(Q_{ij}=4)+\beta_{5}I(Q_{ij}=5) \}$.}
\end{table}

With $\sigma_{b}=0.5$, the ML approach is almost unbiased for estimating
$p$ as well as the regression coefficients. When $\sigma_{b}$ increases
to 1.5, the ML estimator becomes slightly biased with poor CI coverage,
especially for $p$ and $\beta_{0}$. The estimation of other parameters
appears to be robust to the model misspecification, except that the
SEs for $\beta_{2}$ and $\beta_{3}$ overestimate the true variability.
On the other hand, the five GEE methods all perform quite well, in
terms of little bias and close-to-nominal coverage rates. Similar
to the previous simulation study, we observed that about 1\% of the
GEE simulations did not converge for $\sigma_{b}=1.5$. Although the
maximum-likelihood approach is biased, its standard error is much
smaller than the GEE approaches. For example, the ML estimator for
$p$ in Table~\ref{tab4} has a SE only a quarter as large as that
of the GEE-CI and GEE-CE estimators. As a result of the variance-bias
trade-off, the mean squared error for the ML estimator is still smaller
than GEE. If the interest is in estimation, one can still argue that
the ML performs better; but if the interest is in hypothesis testing,
GEE methods are preferred, as they are more robust and preserve the
correct Type~I error rate.

From the above two sets of simulation studies, we would generally
recommend the ML estimator in practice because of its high efficiency.
The correlation structure of the outcome is critical in identifying
the zero-inflation process. Therefore, a full parametric assumption
for the correlation can lead to good efficiency in the estimation.
However, if this parametric assumption does not hold, the ML estimator
could have poor CI coverage rates. In order to perform hypothesis
testing, we would prefer the GEE approaches, which only rely on the
correct mean model and are not sensitive to the working correlation
assumption. Among the five GEE approaches, the GEE-CI and GEE-CE are
the most favorable, because they are more efficient by exploiting
the dependence structure induced by the zero-inflation process. In
practice, the GEE-CI and GEE-CE estimators may be computed in conjunction
with the ML estimator as a sensitivity analysis.

\section{Dating violence data example}\label{sec5}

In this section we fit our proposed ML and GEE models to the dating
violence example, together with the naive GLMM model. A total of 2787
students were enrolled in the study, among which 664 left all the
dating violence questions blank. These 664 subjects are either not
in a relationship, and thus skipped these questions, or they did not
respond at all to the whole survey. In the remaining 2123 subjects,
39 were excluded because they only answered part of the dating violence
questions, and 61 were excluded because they have missing data in
the covariates. The final analysis sample was $N=2023$. The clustered
outcomes of interest ($Y_{ij}$) are the ten questions of dating
violence, including
five victimization and five perpetration questions. We can see from
Figure~\ref{fig1} that the frequency histogram of ``yes'' responses
shows a huge spike at 0. It seems likely that some students who answered
all the questions with ``no'' were not in a relationship,
that is, a zero inflation of the outcome. Define the latent variable
$Z_{i}=1$ if the subject is in a relationship and 0 otherwise. We
included gender ($\mathrm{GENDER}_{i}$), depressive symptoms ($\mathrm{DS}_{i}$), family
relationship ($\mathrm{FR}_{i}$) and family influence ($FI_{i}$) as the
predictors of $Y_{ij}$ given $Z_{i}=1$. The DS score comes from
the questionnaire of depressive symptoms and is on the continuous
scale ranging from 1 to 5, with the larger score indicating worse
depressive symptoms. The FR (ranging from 0 to 10) measures the participant's
satisfaction with the relationship in his/her family, with 10
being a very good relationship. The FI score (ranging from 1 to 7) is
the family influence on the participants not verbally or physically
abusing their romantic partner, with a higher score being greater influence.
We adjust for question number as a factor and question type (victimization
vs. perpetration), in order to account for different prevalence
of yes responses. The interactions between question type and other
covariates ($\operatorname{GENDER}_{i}$, $\operatorname{DS}_{i}$, $\operatorname{FR}_{i}$ and $\operatorname{FI}_{i}$) are also
included. The summary statistics of these variables are described
in Table~\ref{tab5}.

%
\begin{table}
\tabcolsep=0pt
\caption{Summary statistics of the dating violence example. Percentage is
reported for the categorical variables and mean (standard deviation) is
reported for the continuous variables}\label{tab5}
\begin{tabular*}{\tablewidth}{@{\extracolsep{\fill}}@{}ld{2.6}@{}}
\hline
\textbf{Variable} & \multicolumn{1}{c@{}}{\textbf{Summary statistics ($\bolds{n=2023}$)}}\\
\hline
Gender: female & 56.6\%\\
DS score & 2.0~(1.0)\\
Family relationship & 7.4~(2.3)\\
Family influence & 5.7~(1.8)
\\[3pt]
Question 1V---Insult you & 18.5\%\\
Question 1P---Insult your boyfriend/girlfriend & 16.8\%\\
Question 2V---Swear at you & 31.3\%\\
Question 2P---Swear at your boyfriend/girlfriend & 26.1\%\\
Question 3V---Threaten you & 7.2\%\\
Question 3P---Threaten your boyfriend/girlfriend & 5.6\%\\
Question 4V---Push you & 13.5\%\\
Question 4P---Push your boyfriend/girlfriend & 9.9\%\\
Question 5V---Throw object at you & 4.5\%\\
Question 5P---Throw object at your boyfriend/girlfriend & 3.8\%
\\
\hline
\end{tabular*}
\end{table}

Denote $\mathbf{X}_{ij}$ to be the design matrix including all
the covariates and interaction terms mentioned above. We fit the probit
random effects model
%
\begin{equation}
P(Y_{ij}=1\mid\mathbf{X}_{ij},b_{i},Z_{i}=1)=
\Phi\bigl(\mathbf{X}_{ij}'\bolds{\gamma}+b_{i} \bigr),\label{eq3-cond}
\end{equation}
where $b_{i}\sim N(0,\sigma_{b}^{2})$ is the random intercept. This
model has the same marginal mean as the marginal probit regression
model:
%
\begin{equation}
P(Y_{ij}=1\mid\mathbf{X}_{ij},Z_{i}=1)=\Phi
\bigl(\mathbf{X}_{ij}'\bolds{\beta}
\bigr)\label{eq4-marg}
\end{equation}
as $\bolds{\beta}=\frac{\bolds{\gamma}}{\sqrt{1+\sigma_{b}^{2}}}$
for $k=0,\ldots,6$. For comparative purposes, we report the marginal
regression coefficients $\bolds{\beta}$ for all the analyses.

%
\begin{sidewaystable}
\tabcolsep=0pt
\tablewidth=\textwidth
\caption{Parameter estimation and standard errors using the ML, GLMM
and GEE methods for the dating violence victimization example}\label{tab6}
\begin{tabular*}{\tablewidth}{@{\extracolsep{\fill}}@{}ld{2.10}d{2.10}d{2.10}d{2.10}d{2.10}d{2.10}d{2.10}@{}}
\hline
\textbf{Parameter} & \multicolumn{1}{c}{\textbf{ML}} &
\multicolumn{1}{c}{\textbf{GLMM}} & \multicolumn{1}{c}{\textbf{GEE-MI}} &
\multicolumn{1}{c}{\textbf{GEE-CI}} & \multicolumn{1}{c}{\textbf{GEE-ME}} &
\multicolumn{1}{c}{\textbf{GEE-CE}} & \multicolumn{1}{c@{}}{\textbf{GEE-UN}}\\
\hline
(Intercept) & -0.192~(0.176) & -0.746~(0.177) & -0.055~(0.254) & -0.159~(0.225) & -0.497~(0.222) & -0.341~(0.221) & -0.201~(0.232)\\
Question 2 & 0.528~(0.037) & 0.374~(0.024) & 0.578~(0.071) & 0.508~(0.057) & 0.475~(0.058) & 0.488~(0.057) & 0.529~(0.061)\\
Question 3 & -0.754~(0.043) & -0.577~(0.031) & -0.814~(0.069) & -0.750~(0.061) & -0.716~(0.063) & -0.729~(0.061) & -0.768~(0.063)\\
Question 4 & -0.332~(0.035) & -0.250~(0.026) & -0.360~(0.053) & -0.335~(0.048) & -0.315~(0.047) & -0.324~(0.047) & -0.339~(0.049)\\
Question 5 & -1.003~(0.050) & -0.773~(0.036) & -1.069~(0.083) & -0.996~(0.073) & -0.949~(0.077) & -0.969~(0.074) & -1.013~(0.075)\\
Question type & 0.222~(0.122) & 0.158~(0.089) & 0.264~(0.173) & 0.234~(0.150) & 0.215~(0.146) & 0.223~(0.145) & 0.215~(0.155)\\
DS score & 0.192~(0.036) & 0.178~(0.035) & 0.250~(0.043) & 0.220~(0.038) & 0.226~(0.037) & 0.219~(0.037) & 0.230~(0.039)\\
Gender & 0.132~(0.067) & 0.153~(0.054) & 0.171~(0.076) & 0.118~(0.068) & 0.186~(0.067) & 0.160~(0.066) & 0.164~(0.070)\\
Family relationship & -0.037~(0.014) & -0.035~(0.012) & -0.042~(0.019) & -0.039~(0.017) & -0.041~(0.015) & -0.041~(0.016) & -0.041~(0.017)\\
Family influence & -0.114~(0.017) & -0.087~(0.015) & -0.137~(0.021) & -0.135~(0.018) & -0.095~(0.017) & -0.112~(0.017) & -0.123~(0.018)\\
Question type${}\times{}$DS score & 0.037~(0.025) & 0.026~(0.018) & 0.034~(0.032) & 0.037~(0.029) & 0.028~(0.026) & 0.034~(0.028) & 0.035~(0.029)\\
Question type${}\times{}$gender & -0.409~(0.051) & -0.300~(0.037) & -0.446~(0.066) & -0.392~(0.057) & -0.373~(0.056) & -0.379~(0.056) & -0.399~(0.059)\\
Question type${}\times{}$family & -0.003~(0.010) & -0.002~(0.008) & -0.009~(0.014) & -0.007~(0.012) & -0.006~(0.012) & -0.006~(0.012) & -0.005~(0.013) \\
\quad relationship \\
Question type${}\times{}$family& 0.025~(0.012) & 0.019~(0.009) & 0.031~(0.015) & 0.025~(0.013) & 0.026~(0.013) & 0.024~(0.013) & 0.028~(0.013)\\
\quad influence
\\[3pt]
$\Pr(Z=1)$ & 0.571~(0.031) & 1 & 0.523~(0.062) & 0.626~(0.075) & 0.690~(0.104) & 0.663~(0.089) & 0.587~(0.070) \\
$\sigma_{b}$ & 1.033~(0.065) & 1.627~(0.055) & \multicolumn{1}{c}{--} & \multicolumn{1}{c}{--} & \multicolumn{1}{c}{--} & \multicolumn{1}{c}{--} &\multicolumn{1}{c@{}}{--} \\
\hline
\end{tabular*}
\end{sidewaystable}

The results of the ML, GLMM and GEE estimations are listed in Table~\ref{tab6}. From the ML estimation, we can see that all four subject-level
covariates are significant: the probability of dating violence perpetration
was higher for females, those who are more depressed, those who
have a worse relationship in their family, and those who are less
influenced by their guardians. The interaction terms between question
type and gender, and question type and family influence were both
significant, suggesting (a)~boys are more likely to be the victims
of dating violence, and (b)~the family influence has a slightly higher
impact on dating violence perpetration than victimization. The finding regarding
greater male dating violence victimization in this age is in line with
previous studies [\citet{Fos96}; \citet{Arc00}]. The impacts
of depression score and family relationship are similar regardless
of question type. The directions of association are expected and consistent
with some of the findings in \citet{Hayetal13}. The zero-inflation
probability is estimated to be 0.571, that is, we expect about 43\%
of the sample (about 860 subjects) to be structural zeros. We suspect
that a majority of
them were not in a relationship, but
answered all the dating violence questions with ``no.''
However, there may be alternative reasons for the zero inflation, for example,
some kids may give socially desirable answers in the survey and hence
underreport dating violence.
However, we believe that this only accounts for a small fraction of the
structural zeros.
The likelihood ratio test statistic for zero inflation is $\Lambda
=65.2$ ($p$-value${}<{}$0.001).
The parameter estimations by GEE
are generally close to ML, but the standard errors are larger. The
naive GLMM method estimated smaller covariate effects, which could
be biased due to ignoring the zero-inflated nature of the data.

As pointed out by a referee, the zero-inflation problem could be
avoided by including
a filter question of asking whether the subject had a relationship or
not. The filter
question was not included because the study investigators felt that it
was an unreliable question to ask. Relationships between teenagers
today cannot easily be characterized, and the investigators felt that
explicitly asking this question may limit important responses about
violence [\citet{Shoetal13}]. There are other cases where the
susceptible population cannot be ascertained accurately. For example,
in drug abuse studies, specifically asking whether individuals are drug
abusers might not be a question that would result in a reliable
response. But we may be interested in whether the abusers seek
particular types of treatment. Additionally, this approach may be
relevant to questions regarding immigration status, where
there may be legal implications (perceived or real) in answering, and
in questions on mental health
status, where the respondent may have a reduced ability to accurately
report their status.

\section{Discussion}\label{sec6}

In public health research, excessive zero responses may occur if the
population which is susceptible to respond is not carefully screened or
is unknown. The resulting zero inflation may have an effect on the
results obtained by conventional methods of analysis. In the NEXT
Generation Health Study, investigators were interested in identifying
predictors of dating violence in teenagers. Examining these regression
relationships are of interest for those individuals who are in a
relationship (i.e., the susceptible condition). Many more individuals
completed this study component than investigators thought would be in a
relationship at this age. This led to what appears to be zero-inflated
clustered binary data. We developed both ML and GEE approaches for
analyzing such data. Through simulations and analysis of the real data
example, we found that the ML approach is substantially more efficient
than the GEE approaches. However, under moderate model
misspecification, the ML approach may result in biased inference. It is
recommended that, as a sensitivity analysis, both ML and GEE approaches
be applied in applications.

In the GEE approach, we treat the regression parameters and nuisance
working correlation as orthogonal, that is, a GEE1 approach with the
parameters in the working correlation estimated by a moment estimator.
Potential efficiency gain could be achieved using an improved version
of GEE1 [\citet{Pre88}] or GEE2 [Prentice and Zhao (\citeyear{PZ91});
Liang, Zeger and Qaqish (\citeyear{LZQ92})], by establishing another set of estimating
equations on the second moment. However, GEE2 requires a correct
variance structure with working third and fourth moment model, which is
hard to verify with the presence of zero inflation, and results in bias
under second, third and fourth moment misspecification. The previous
work of \citet{HalZha04} adopted Prentice and Zhao's GEE2 that
maintains the parameter orthogonality in their second moment estimating
equations, and they argued that only making a first moment assumption
may lead to parameter nonidentifiability. This is true in their case,
where the zero-inflation probability is at the observation level, so
$p_{ij}$ and $\mu_{ij}^{Z}$ might be confounded in $\mu_{ij}^{M}$.
However, in the case of subject-level zero inflation, we have shown the
model identifiability of the GEE1 estimators.

In principle, zero-inflated models cannot be identified
nonparametrically; parametric assumptions for the nonzero part play a
fundamental role in model estimation. For example, zero inflation in
Poisson and binomial data can be determined by the lack of fit in the
zero cells of these respective distributions. In this paper, we assume
that the nonzero distribution is given by a generalized linear mixed
model with normal random effects. The ML approach exploits the
correlation structure in order to distinguish structural zeros and
random zeros. Intrinsically, GEE only uses the mean structure of the
binary data in order to estimate regression parameters and, unlike ML,
does not use the entire distribution for estimation.


In our application, there were very few missing data. However, in many
studies with sensitive psychological or behavioral questions, there may
be informative missingness. An advantage of the ML approach is that it
can more easily be extended to account for informative missing
responses [see \citet{FolWu95}, e.g.]. The proposed methodology
implicitly assumes that the subjects answer the questions truthfully.
If this assumption does not hold, the parameter estimation is likely to
be biased. We could formulate a likelihood approach if we had good
prior information about the distribution of false negative occurrence
across the questions. This would require a verification subsample
corresponding to the survey questions (maybe obtained through
interviews of parents and friends) on a fraction of teenagers.

The zero-inflation probability in our model is assumed to be constant,
but it is straightforward to include covariates in both ML and GEE
approaches. In addition, our focus is on a cross-sectional inference,
that is, analyzing the dating violence data at one point in time (11th
grade). Understanding behavior change from adolescence over time is
interesting but also challenging. In the longitudinal setting, the
zero-inflation probability is time-dependent that can probably be
modeled by a latent Markov process. We will leave it for future exploration.

\section*{Acknowledgments}
We greatly appreciate and enjoyed reading the very insightful comments
and suggestions by the Associate Editor and two anonymous referees.
Their comments not only remarkably improved the quality of the paper,
but also motivated our deeper thoughts of possible extensions of the
methodology in the future.

This study utilized the high-performance computational capabilities of
the Biowulf Linux cluster at the National Institutes of Health,
Bethesda, MD (\surl{http://biowulf.nih.gov}).


\begin{supplement}[id=suppA]
\stitle{Supplement to
``Mixed model and estimating equation approaches for zero
inflation in clustered binary response data with application to a
dating violence study''}
\slink[doi]{10.1214/14-AOAS791SUPP} 
\sdatatype{.pdf}
\sfilename{AOAS791\_supp.pdf}
\sdescription{Supplement~A: Additional simulation one.
Examine the performance of the proposed model with a smaller sample size ($N=500$).
Supplement B: Additional Simulation Two.
Examine the sensitivity
of assuming a constant zero-inflation probability when the probability
is affected by covariates.
Supplement C: Additional Simulation Three.
Examine the performance of zero-inflated beta-binomial
model.}
\end{supplement}

%

\printaddresses
\end{document}